\documentclass[referee]{raa}
\usepackage{graphicx,times}
\usepackage{natbib}
\usepackage{amssymb,amsmath}
\bibpunct{(}{)}{;}{a}{}{,}

\usepackage[a4paper=true,dvipdfm=true,pagebackref=true]{hyperref}
\hypersetup{pdftitle = The title of my PDF, pdfauthor = My name, pdfsubject= The subject, pdfkeywords = keyword1 keyword2 keyword3}
\hypersetup{colorlinks = true, linkcolor = green, anchorcolor = red, citecolor = blue, filecolor = red, pagecolor = red, urlcolor = red}

\begin{document}

   \title{Broken-up Spectra of the Loop-top Hard X-ray Source during a Solar Limb Flare}

 \volnopage{ {\bf 20XX} Vol.\ {\bf X} No. {\bf XX}, 000--000}
   \setcounter{page}{1}

   \author{Hao Ning\inst{1}, Yao Chen\inst{1}, Jeongwoo Lee\inst{1}, Zhao Wu\inst{1}, Yang Su\inst{2}, Xiang-Liang Kong\inst{1}
   }

   \institute{ Shandong Provincial Key Laboratory of Optical Astronomy and Solar-Terrestrial Environment, and Institute of Space Sciences, Shandong University, Weihai, Shandong 264209, China; {\it yaochen@sdu.edu.cn}\\
        \and
        Purple Mountain Observatory, Chinese Academy of Sciences, Nanjing 210008, China\\
\vs \no
   {\small Received 20XX Month Day; accepted 20XX Month Day}
}

\abstract{Solar hard X-rays (HXRs) appear in the form of either footpoint sources or coronal sources, and each individual source provides its own critical information on acceleration of nonthermal electrons and plasma heating. Earlier studies found that the HXR emission in some events manifests a broken-up power-law spectrum with the break energy around a few hundred keV based on spatially-integrated spectral analysis, without distinguishing the contributions from individual sources. In this paper, we report the broken-up spectra of a coronal source studied using HXR data recorded by \textit{Ramaty High Energy Solar Spectroscopic Imager} (\textit{RHESSI}) during the SOL2017-09-10T16:06 (\textit{GOES} class X8.2) flare. The flare occurred behind the western limb with its foot-point sources mostly occulted by the disk, and we could clearly identify such broken-up spectra pertaining solely to the coronal source during the flare peak time and after. Since a significant pileup effect on the \textit{RHESSI} spectra is expected for this intense solar flare, we have selected the pileup correction factor, $p = 2$. In this case, we found the resulting \textit{RHESSI} temperature ($\sim$30 MK) similar to the GOES soft X-ray temperature and break energies of 45--60 keV. Above the break energy the spectrum hardens with time from spectral index of 3.4 to 2.7, and the difference of spectral indices below and above the break energy increases from 1.5 to 5 with time.
We, however, note that when $p = 2$ is assumed, a single power-law fitting is also possible with the \textit{RHESSI} temperature higher than the GOES temperature by $\sim$10 MK. Possible scenarios for the broken-up spectra of the loop-top HXR source are briefly discussed.
}

   \authorrunning{H. Ning et al. }            
   \titlerunning{Broken-up spectra of loop-top source}  
   \maketitle

%
\section{Introduction}           
\label{sec1}
In the standard picture of solar flares called CSHKP model \citep{carm64,stur66,hirayama74,kopp76}, energetic particles are accelerated through magnetic reconnection in the corona, and those moving downward along the field lines collide with the dense chromosphere to emit hard X-rays (HXRs) and $\gamma$-rays via bremsstrahlung process. This produces intense foot-point HXR sources and heats the chromospheric plasmas to $\sim$10~MK. In a follow-up process called chromospheric evaporation, the heated plasmas then move upward to fill in the post-flare loops to emit soft X-rays (SXRs).

In addition to the HXR footpoint sources and the SXR loop source, \cite{masuda94} found another HXR source located above the SXR loop, called the above-the-loop-top source or ``Masuda-type'' source. Similar sources have been observed in other events \citep[e.g.,][]{k10cs,oka15}, and they are considered as an important clue to particle acceleration in the corona.

A coronal HXR source is expected to be more hardly visible because plasma is more dilute in the corona and bremsstrahlung emission there is weaker than in the footpoints. Also considering the limited dynamic range of present HXR imaging instruments, coronal sources are better observed in limb events with footpoint sources partially or fully occulted. Statistical surveys of limb flares found that coronal sources with significant non-thermal components form quite frequently \citep[see, e.g.,][]{k08cs,effenberger17}.
In some events, coronal sources are found to be associated with very-dense loop tops and weak footpoint emission, and may emit in thick-target bremsstrahlung \citep[e.g.,][]{sui04,veronig04}.

HXR spectra of solar flares usually appear in two components: thermal component at low energy range ($<$10--20~keV) and non-thermal component with a power law at higher energy range. In terms of temporal variation of HXR spectra, the majority of HXR spectra exhibit soft-hard-soft (SHS) evolution \citep[e.g.,][]{parks69,grigs04}, while a small fraction of events manifests a soft-hard-harder (SHH) behavior \citep[see, e.g.,][]{frost71,cliver86}. These temporal variations have been used as constraints in modeling acceleration and transport of energetic electrons.

The property of thermal components of HXR spectra varies in different events. Usually, the spectra contain a hot component with a temperature of $\sim$10~MK and high emission measure. In some cases, however, a so-called super-hot component has been reported \citep[e.g.,][]{caspi10,caspi14,ning18}, with a temperature of over 30~MK, while the EM value is relatively lower. Plasmas with such high temperatures are believed to be a result of the direct heating in the corona, rather than the chromospheric evaporation process.

In addition to the thermal/nonthermal components of solar HXR spectra, the nonthermal spectra may appear broken upward at high energies, namely, spectral hardening at higher energies \citep[e.g.,][]{suri75,share03,shih09,ackermann12}. According to the statistical study by \cite{kong13}, the break energy, $E_b$, is usually around a few hundred keV, and the difference of spectral indices below and above $E_b$ ranges from 1.5 to 2.5.

Two scenarios have been proposed for explaining the broken-up spectra. In one scenario, the broken-up spectra is attributed to the superposition of multiple sources with different spectral indices \citep{k08hardening}. In the other scenario, it is attributed to the intrinsic property of the acceleration
mechanisms such as the diffusive acceleration by the flare termination shock \citep{lig13} or the stochastic acceleration by wave-particle interactions \citep{H92}, which results in energetic electrons with a broken-up energy spectrum. It is also suggested that such broken-up spectra are due to the trap-and-precipitation of electrons within the loops \citep{park97,lee00,minoshima08}.

To explore the origin of broken-up spectrum of HXRs, it will be crucial to disentangle the contributions from spatially distinct sources, since most previous studies are based on spatially-integrated HXR spectrum. In this paper, we present our study of a spatially-resolved limb flare that exhibits significant broken-up spectrum in the loop-top HXR source using imaging spectroscopy with \textit{Ramaty High Energy Solar Spectroscopic Imager} \citep[\textit{RHESSI};][]{huford02,lin02}. Section~\ref{sec2} presents the data and an overview of the event. Section~\ref{sec3} shows the analysis of the \textit{RHESSI} data of this event. The discussion and the conclusions are presented in the last two sections.

\section{Observational Data and Event Overview}\label{sec2}

We use the data from \textit{RHESSI} and Atmosphere Imaging Assembly \citep[AIA;][]{lemen12} onboard the \textit{Solar Dynamics Observatory} \citep[\textit{SDO};][]{pesnell12}. \textit{RHESSI} detects X-ray and $\gamma$-ray sources in the Sun with high cadence (4~s), and spatial (3{\arcsec}) and energy resolutions (as high as 1~keV). AIA has the capability of imaging plasma structures at temperatures from 20,000~K to over 20~MK, with high spatial (0.6{\arcsec} pixel size) and temporal (12~s) resolutions in 10 different UV and EUV passbands. AIA images at 171~($\sim$0.6~MK) and 193~\AA{} ($\sim$1.6 and 18~MK) are mainly used.

\begin{figure}[!h]
\centering
\includegraphics[width=0.9\textwidth]{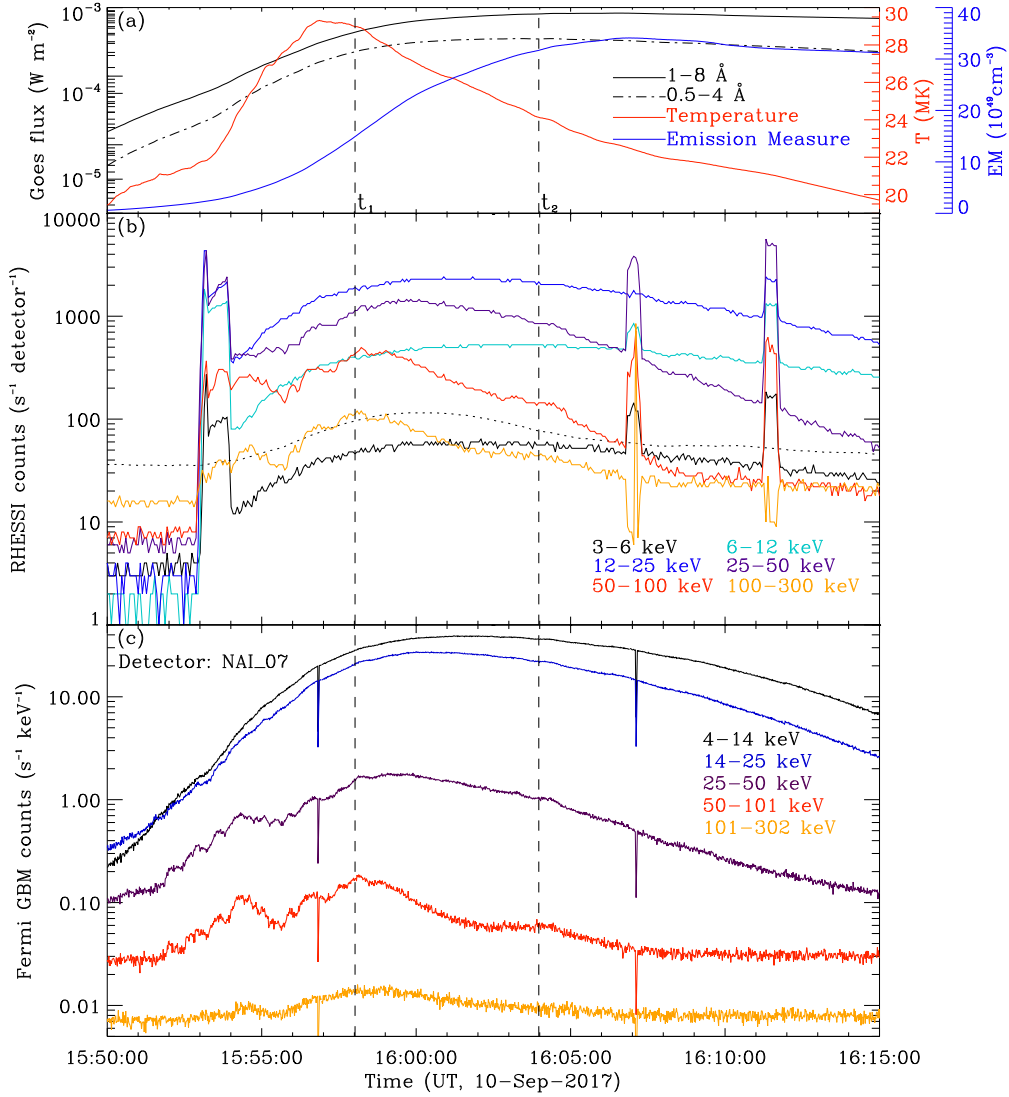}
\caption{Lightcurves of the 2017-Sep-10 flare observed by \textit{GOES} (a), \textit{RHESSI} (b), and \textit{Fermi} GBM (c). The temporal profiles of the temperature (T) and emission measure (EM) derived from the \textit{GOES} data are also shown in panel (a). In panel (b), the dotted line shows the temporal variation of the background of \textit{RHESSI} data. The two vertical dashed lines
mark the peak time of HXR ($t_1$, 15:58~UT) and SXR ($t_2$,
16:04~UT), respectively.}\label{fig1}
\end{figure}

The \textit{GOES} X8.2-class flare of the study occurred on 2017 September 10 from the active region (AR) 12673 at the western limb of the solar disk, accompanied by a fast CME. There are 4 X-class flares and 27 M-class flares released from this AR \citep[e.g.,][]{yang17,sharykin18}. During the event of this study, part of the active region is on the back side of the solar disk, so that the flare footpoints are partially occulted. This yields a nice observation of both the morphology of the flaring structures with AIA and the coronal X-ray sources with \textit{RHESSI}.

The X-ray lightcurves observed by \textit{GOES}, \textit{RHESSI}, and \textit{Fermi} Gamma-ray Burst Monitor \citep[GBM,][]{Meegan09} are shown in Figure~\ref{fig1} (a), (b), and (c). \textit{RHESSI} was at the orbit night until 15:52~UT, after which the X-ray flux at 3--300~keV rises sharply. From 15:52~UT to 15:54~UT, the attenuator state changed from A0 to A3 to cause the large jumps of the X-ray flux, and the two abnormal peaks after 16:04~UT are also caused by the attenuator state change (from A3 to A1). We thus focus on the data from 15:54~UT to 16:04~UT. The X-ray flux of 100--300~keV arises to the peak at around 15:58~UT ($t_1$), and then decreases gradually.

 At high energies (50--300~keV), we can see that the time profiles of the X-ray flux obtained by \textit{Fermi} GBM are similar to those of \textit{RHESSI}. Therefore, here in this paper we focus on the analysis of \textit{RHESSI} data. Note that the GBM data presented here are from the detector NAI\_07, which is not the most sunward-facing one during this flare. This is to avoid the strong pileup and saturation issues since the flare is extremely intense. At lower energy, the flux peaks later ($\sim$16:00~UT). The \textit{GOES} SXR flux starts to increase earlier ($\sim$15:50~UT) with a more gradual profile, and reaches its peak at 16:04~UT ($t_2$).

\begin{figure}[!h]
\centering
\includegraphics[width=0.9\textwidth]{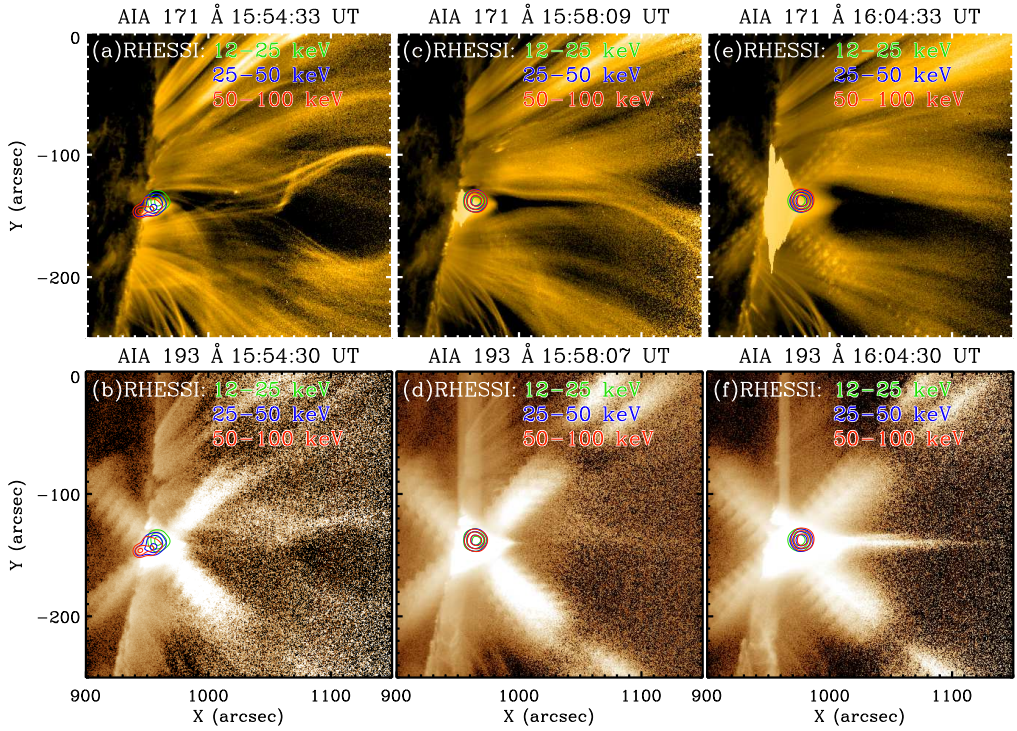}
\caption{EUV images of the event observed by \textit{SDO}/AIA and
\textit{RHESSI} HXR sources. Panels (a), (c), and (e) are images
at 171~\AA{}, and panels (b), (d), and (f) show images at
193~\AA{}. The contours represent the \textit{RHESSI} sources at
12--25 keV (green), 25--50 keV (blue) , and 50--100 keV (red).
Contour levels are given by 30\%,60\%, and 90\% of the maximal
flux in each energy interval. The three moments of the EUV images correspond to the early stage of impulsive phase, the HXR peak time (t1), and the SXR peak time (t2), respectively. The
HXR images are reconstructed by the CLEAN algorithm using
detectors 3, 6, and 8. The time intervals for reconstructing the
HXR sources are 15:54:15~UT--15:55:15~UT,
15:57:30~UT--15:58:58~UT, and 16:04:00~UT--16:05:00~UT.}\label{fig2}
\end{figure}

Figure~\ref{fig2} shows the EUV images observed by AIA. Initially, a bright bubble-like structure is observed at 171 and 193~\AA{}, which ejects outward as the CME propagating at a speed of 400--600~km~s$^{-1}$. During the process, the loop structures are stretched outward and re-close according to the data at 171~\AA{}. A bright ray structure appears in 193~\AA{} images. Other studies on this event suggest that the bubble represents a flux rope structure and the bright ray corresponds to the large-scale current sheet \citep[see, e.g.,][]{cheng18}. Other aspects that have been investigated are the complete eruptive process and characteristics of reconnection inflows \citep{yan18}, the association of the eruption of the bubble-like structure and flare energy release \citep{long18}, spatial distribution of high-energy electrons in a large region from the microwave perspective \citep{Gary18}, and turbulent features within the ray-like structure \citep{cheng18}.

\section{Imaging and Spectral Analysis of \textit{RHESSI} data} \label{sec3}

We focus on spectral characteristics of HXR emission from the loop top source, during and after the peaking time of the X-ray flux. As Figure~\ref{fig2} shows, three time intervals are selected to represent the rising impulsive phase (panels a--b), the HXR peak ($t_1$, panels c--d), and the SXR peak ($t_2$, panels e--f). The sources are shown, superimposed onto the EUV images taken at the nearby times.

At around 15:54~UT, the centroid of the 12--25~keV source is located in the corona, around 10{\arcsec} above the limb, co-spatial with the loop top observed at 171~\AA{}. The source at 25--50~keV can be divided into two components with one coronal source and one foot-point source. The coronal component is slightly lower than the 12--25~keV loop-top source by about 4--5{\arcsec}, while the foot-point component is relatively weak with an intensity of about 30\% of the maximal flux. The 50--100~keV source also has one footpoint and one coronal components with the footpoint one being more intense.

At the HXR peak time ($t_1$), the centroids of different energy band sources are basically co-spatial with each other and located in the bright loop top observed at EUVs around 15\arcsec{} to 20\arcsec{} above the disk. At the \textit{GOES}-SXR peak time ($t_2$), the sources rise with the expansion of post-flare loops, and reach up to $\sim$30\arcsec{} above the limb. All sources are very close to each other, yet there is a trend that the more energetic source, the higher it is located.

The location of this HXR source above the limb has an advantage in that strong footpoint emissions are occulted, allowing us to focus on the spectral property of the looptop source. As a major difficulty, however, a significant pileup effect on the \textit{RHESSI} spectra is expected, since this flare is so intense to acquire the \textit{GOES} class X8.2. We therefore investigate the pileup effect in the spatially integrated \textit{RHESSI} spectra, after which we proceed to study the temporal evolution and imaging spectroscopy.

\subsection{Effect of Moderate pileup on the HXR spectra}

\begin{figure}[!h]
\centering
\includegraphics[width=0.9\textwidth]{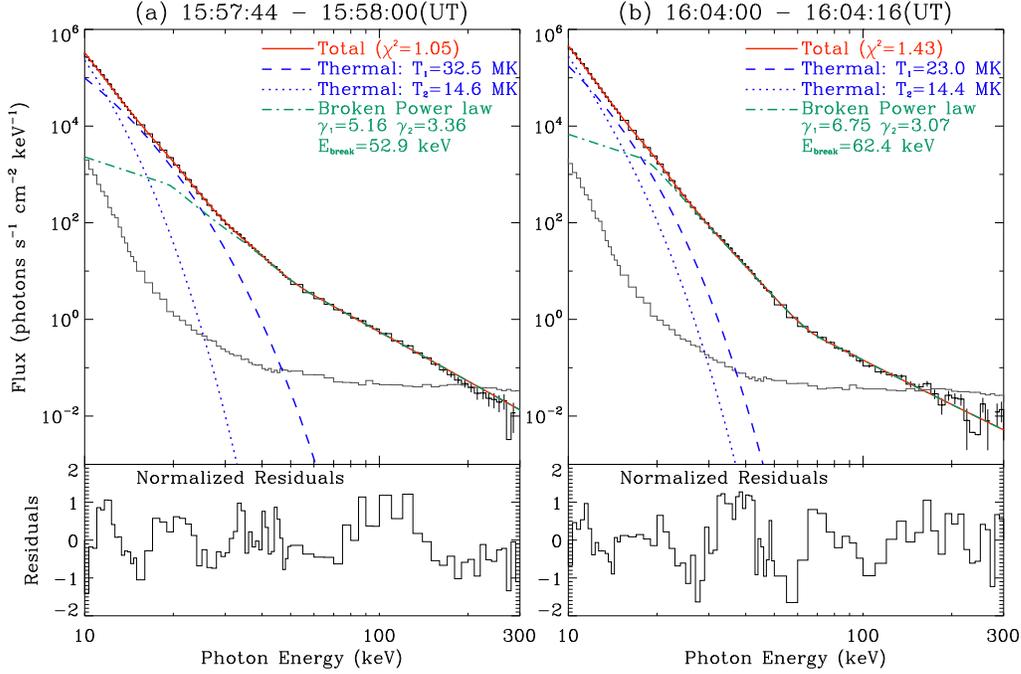}
\caption{Spectral analysis results of the spatially-integrated
\textit{RHESSI} flux spectra. (a) Spectrum at $t_1$. (b) Spectrum
observed at $t_2$. The histograms with black error bars are the
background-subtracted spectra, and the red lines represent the
total fitting results. The background is given in grey. Blue dotted lines represent the hot
components, and blue dashed lines represent super-hot components.
Non-thermal components are given as green lines, representing
broken power-law. The normalized residuals are shown as the
histograms in the bottom-most panels. Data from detectors 1, 3, 6,
and 8 are used.} \label{fig3}
\end{figure}

We start the spectral fitting of the spatially-integrated \textit{RHESSI} spectra with default parameters for the pileup correction. $\footnote{Two photons close in time are detected as one photon and have their energies added. Pileup of three or more photons is possible, but at a much lower probability \citep{Smith02}.}$ Within the SSW, this is the case when the pileup coefficient, $p$, is set to 1.
The spectral fitting results around $t_1$ and $t_2$ are shown in Figure~\ref{fig3}. Panel (a) shows the spectra integrated from 15:57:44~UT to 15:58:00~UT, and panel (b) shows the spectra integrated from 16:04:00~UT to 16:04:16~UT. In this period, the footpoint source is not visible, and the loop-top source is dominant in the HXR emission.

The spectrum at $t_1$ can be fitted with two thermal components and a non-thermal component, and $\chi^2$ is 1.05. Below 20--30~keV, the fit is given by a hot component with a temperature of 14.6~MK and a super-hot component with a temperature of 32.5~MK. We obtain the emission measure defined by $\text{EM}= n^2 V$ where $n$ is the plasma density and $V$ is the source volume. The EM of the two components are $2.9\times10^{51}$ and $3.6\times10^{49}$~cm$^{-3}$, respectively. The spectrum of higher energy range can be fitted with a broken power-law distribution. The $E_b$ is found to be 52.9~keV; the spectral index below $E_b$ is $\gamma_1=5.16$ and the spectral index above $E_b$ is $\gamma_2=3.36 $. In summary, a significant broken-up feature of the spectra above $E_b$ is observed, and the difference between the two spectral indices is $\Delta \gamma \equiv \gamma_1-\gamma_2 = 1.8$.

Similarly, the \textit{RHESSI} spectrum at $t_2$ (Figure~\ref{fig3} (b)) could also be well-fitted by two thermal components (hot and super-hot) and one broken power-law component, with $\chi^2=1.43$. Due to the low photon count at high energies (above 200~keV), the fitting uncertainty is large there. We however focus on the spectra below 200~keV. In comparison with the spectrum at $t_1$, the broken-up feature of the spectrum at $t_2$ becomes more significant with $E_b = 62.4$~keV, $\gamma_1 = 6.75$, and $\gamma_2 = 3.07$. Thus, $\Delta\gamma = 3.68$, larger than that of $t_1$. In addition, the temperature of the hot component is 14.4~MK, and the temperature of the super-hot component is 23~MK. The EM of the two components are $9.0\times10^{51}$ and $1.6\times10^{50}$~cm$^{-3}$, respectively. Comparing the fitting results, the temperature of the hot component around $t_2$ is close to that around $t_1$; the temperature of the super-hot components around $t_2$ is lower than that around $t_1$; and EM of both components around $t_2$ are higher than that around $t_1$. The changes of these parameters may indicate the cooling process associated with the super-hot component and the density increase in the post-impulsive stage of the flare.

\subsection{Effect of Strong Pileup on the HXR spectra}

At the time of the event, however, the \textit{RHESSI} detectors suffered from accumulated radiation damage after the long-term operation. This may make the pileup effect more significant, in particular, when observing extreme events such as the present one.
We thus test the possible extent of the pileup effect, by adopting larger values of $p >1$.

\begin{figure}[!h]
\centering
\includegraphics[width=0.9\textwidth]{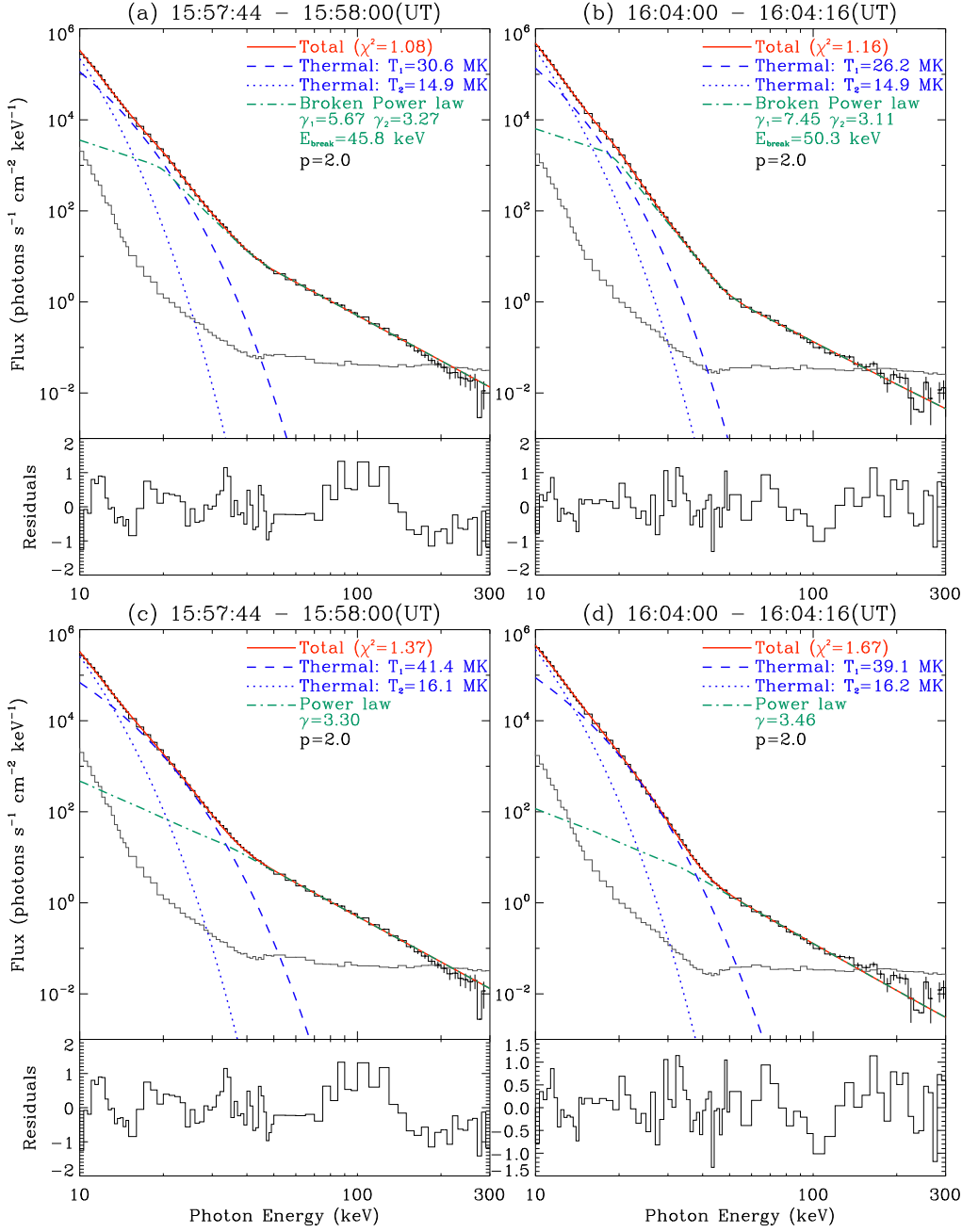}
\caption{Spectral fitting results with different models of the spatially-integrated
\textit{RHESSI} flux spectra. The detectors and the fitting
components are the same as Figure~\ref{fig3}. The coefficient of
pileup effect ($p$) is set to be 2. In panels (a)--(b), the non-thermal part of the spectra is fitted with broken power law, which is fitted with single power law in panels (c)--(d).} \label{fig4}
\end{figure}

In Figure~\ref{fig4}~a--b we plot the fitting results of spatially-integrated spectra at$t_1$ and $t_2$, with $p = 2$. The same models including two thermal components and one broken power-law component are used. We obtained  (a) $E_b = 45.8$~keV, $\gamma_1 = 5.67$, and $\gamma_2 = 3.27$ with $\chi^2 = 1.08$ at $t_1$, and  (b) $E_b = 50.3$~keV, $\gamma_1 = 7.45$, and $\gamma_2 = 3.11$ with $\chi^2 = 1.16$ at $t_2$. The temperatures of the two thermal components are 14.9 and 30.6~MK at $t_1$, while 14.9 and 26.2~MK at $t_2$. Comparing with the default case (Figure 3), we found that the temperature and spectral indices do not change much, but the break energy is lower, for both $t_1$ and $t_2$.
In panels c--d of Figure~\ref{fig4}, we attempt alternative models consisting of two thermal components plus one single power-law component, which becomes plausible when a larger $p$ is adopted. This does not result in much change in the power-law index, $\gamma_2$. However, the temperature increases to $\sim$16~MK and 39--41~MK for the two thermal components, and $\chi^2$ also increases to 1.37 and 1.67, as compared to the default case (Figure~\ref{fig3}).

As shown in Figure~\ref{fig1} (a), the \textit{GOES} data provide temperature around 29~MK at $t_1$, and 24~MK at $t_2$.  This is consistent with the temperature, $T_1$, found for the double power-law spectrum. On the other hand, the single power-law spectrum, which is found to be feasible with a higher $p$, predicts much hotter thermal components reaching $\sim$40 MK. As a comparison, we note that earlier studies also present a super-hot component with such high temperature, and it was indicated that the \textit{RHESSI} temperature can be higher than the \textit{GOES} temperature \citep[e.g.][]{Holman03,caspi10,caspi14,warmuth16}. Therefore the single power-law interpretation cannot be completely ruled out for this flare, considering the significant pileup effect on spectral fitting.

We also tested the spectral fitting with  $p=3$, in which case lower temperatures and harder nonthermal electrons are required, as well as high values of emission measure (EM $\sim 10^{54}~cm^{-3}$). Since this is much higher than the emission measure obtained with $p=1$ or 2 (EM$\sim 10^{51}~cm^{-3}$), we consider it unrealistic. When we enforce a single power-law solution to the spectrum with $p=1$, $\chi^2$ comes out too high ($\sim$3--5) and the temperature gets unrealistically high (over 60~MK). We therefore found the fairly good single power-solution with the pile-up correction around $p= 2$.

\subsection{Temporal Evolution}

\begin{figure}[!h]
\centering
\includegraphics[width=0.9\textwidth]{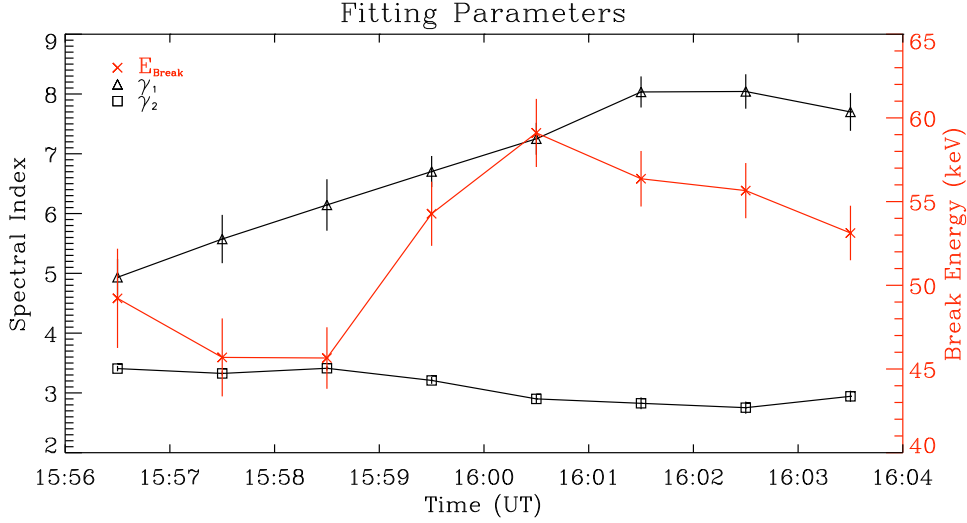}
\caption{Temporal variation of the spectral fit parameters with error bars. Total spectra in eight time intervals from 15:56~UT to 16:04~UT are fitted with two thermal and one broken power-law components, while $p$ is set to be 2. Symbols represent $\gamma_1$ (triangle),
$\gamma_2$ (rectangular), and $E_b$ (cross).} \label{fig5}
\end{figure}

We proceed to investigate the temporal evolution of the HXR spectrum with $p=2$. We divide the period from 15:56~UT to 16:04~UT into 8 intervals with 1-min duration each. We fit the spectrum within each time interval using the above method with two thermal components and one broken power-law component, and found $\chi^2$ around 1 in all cases. The derived temporal profiles of three fitting parameters ($E_b$, $\gamma_1$, and $\gamma_2$) are shown in Figure~\ref{fig5}.

From 15:57~UT to 16:01~UT, $E_b$ increases from 45 to 59~keV, and then declines to around 55~keV from 16:01~UT to 16:04~UT. During the 8 minutes, $\gamma_1$ increases from 5 to 8, and $\gamma_2$ decreases slightly from 3.4 to 2.7, and $\Delta\gamma$ increases continuously from $\sim$1.5 to 5. This indicates that the spectrum below $E_b$ becomes softer and the spectrum above $E_b$ becomes harder, i.e., the broken-up feature becomes more and more significant with time, during the impulsive phase to the decay phase.

\subsection{Imaging spectroscopy}

The spectra analyzed above are spatially integrated over the full disk (hereafter called the total spectra). As mentioned, the foot-point source is not visible after 15:56~UT, therefore, the X-ray spectra analyzed above is unambiguously associated with the loop-top source. To confirm this (and relevant results) using spatially-resolved data, we obtain imaging spectra around $t_1$ (from 15:57:31~UT to 15:58:50~UT) using the SSW OSPEX package. This is done by first reconstructing the HXR sources for different energy bands, and then integrating the photon flux within specified area for each energy band.

\begin{figure}[!h]
\centering
\includegraphics[width=0.9\textwidth]{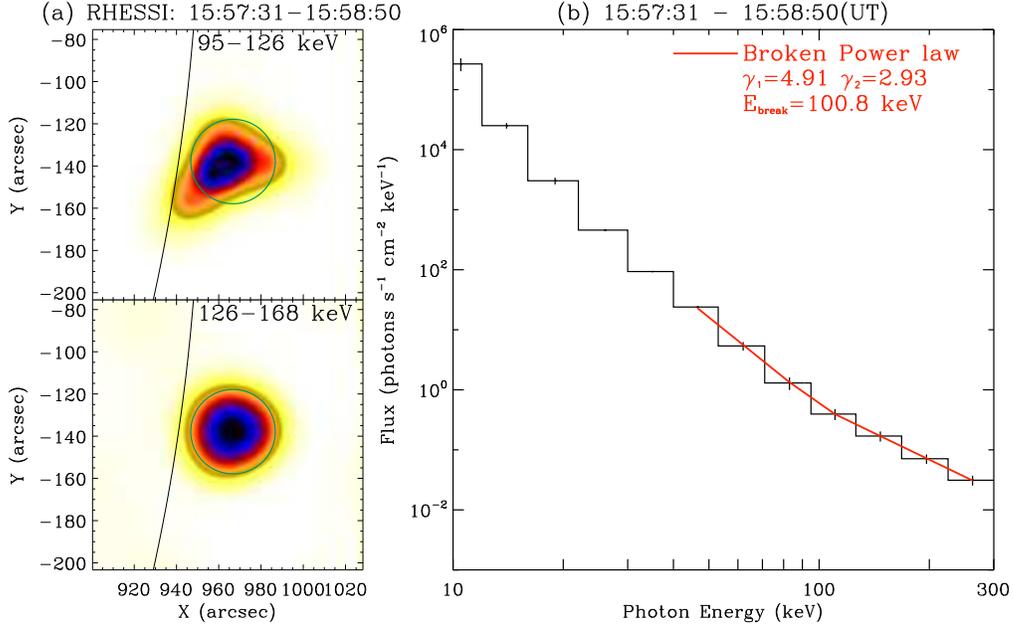}
\caption{Imaging spectroscopic results at $t_1$. (a) Images
reconstructed in the two energy bands. Green circles represent the
region that we used for determining the spectra. (b) The obtained
imaging spectrum. The local spectrum constructed
from the maps is plotted in histogram mode with error bars. Red line is shown as the
broken power-law of 40--300~keV, to fit the non-thermal part of the spectra.
The PIXON algorithm along with the data from detectors 6 and 8 are used to
reconstruct these images.} \label{fig6}
\end{figure}

The loop-top source in the two energy bands (95--126~keV and 126--168~keV) is shown in Figure~\ref{fig6}~(a), from which we calculate the total counts of photon within the green circle for each energy band. The resulting spatially-resolved spectrum is shown as histograms in Figure~\ref{fig6}~(b). The spectrum can be fitted again using two thermal components and one broken-power-law component. Because the energy range is coarse and the pileup effect cannot be corrected, we did not show the total fitting result, but the non-thermal part only (over 40~keV). In this part, the broken-up feature is noticeable, as expected. We found the break energy, $E_b=100.8$~keV, higher than that obtained with the spatially-integrated spectra, and the indices $\gamma_1=4.91$ and $\gamma_2=2.93$ with $\Delta\gamma = 1.98$.

The differences between the spatially-integrated spectroscopy-only result and the result of the imaging spectroscopy are not significant, but may need explanation. We performed the correction of the pile-up effect of \textit{RHESSI} data for the total spectral analysis, but not for the imaging spectroscopic analysis. We also note that wider energy bands are used for the imaging spectroscopy than those for the total spectral analysis in order to have sufficient number of photons. These could have resulted in the different spectral indices and break energy obtained with the two methods.

\section{Discussion}\label{sec4}

\subsection{Nature of the HXR loop-top Source}
In the present event, one footpoint source appears only at the start of the impulsive phase in high-energy passbands (25--100~keV). During the flare peak time and after, the loop-top source is the only HXR source and does not show any significant extension toward the footpoints. It is possible that the footpoint source becomes occulted by the disk due to solar rotation or simply the footpoint source diminished with time.

We here elaborate another possibility that in the loop-top filled up with dense plasma due to chromospheric evaporation, electrons lose most of their energy via thick-target bremsstrahlung, and as a result, not many energetic electrons reach the chromosphere to produce strong footpoint HXR. We calculate the minimum energy, $E_{\rm min}$, of electrons that can reach the footpoint against Coulomb collisions using the \textit{RHESSI} data and geometry of the flaring loop. At $t_1$ the emission measure of the hot component derived from \textit{RHESSI} data is $\rm{EM} \approx 2.9\times 10^{51}$~cm$^{-3}$; the HXR source size is $R\approx 15\arcsec$ (determined with the 10\% contour of the 25--50~keV source at $t_1$); and the distance from the loop-top to the footpoint is estimated as $L\approx$~30{\arcsec}. We thus estimate the density in the loop-top as $n \equiv (3 \rm{EM}/4\pi R^3)^{1/2} \approx 8.0\times10^{11}$~cm$^{-3}$, and the corresponding column depth is $nL \approx 1.6\times10^{21}$~cm$^{-2}$. Thus, under Colulomb collisions, the electrons with energies lower than $E_{\rm min}\approx (8.8~{\rm keV})(nL/10^{19} {\rm cm}^{-2})^{1/2}\approx$ 110~keV cannot reach the footpoint \citep{brown73}. More energetic electrons with $\geq$110~keV can reach the footpoint, but are insufficient in number to produce significant HXR. This supports our thick-target interpretation of the coronal loop-top source and also consistent with some earlier results \citep[see, e.g.,][]{sui04,veronig04,jiang08}.

\subsection{Origin of the Broken-up Spectra}

There are mainly two scenarios for the HXR broken-up spectra, either emitted by two groups of electrons with different power-law indices or from a single group of electrons evolving into a broken power-law distribution in energy (see \S~\ref{sec1}).

The first scenario has been proposed by \cite{k08hardening} for explaining the broken-up feature of the total spectra observed for 3 events, for which they found that the spectra of loop-top sources are harder than the spectra of footpoint sources and that the total spectrum shows hardening. We note that this scenario is consistent with earlier simulations of stochastic acceleration of solar flare electrons by \cite{park97}. In their simulation, the total X-ray emission is regarded as a superposition of thin-target emission of trapped electrons in the loop-top and thick-target emission of precipitating electrons into the footpoint, which explains two flares from the GRS instrument in 1989 and two from the EGRET and BATSE instruments in 1991. \cite{li11} also proposed that the broken-up spectrum can be produced either by the summation of individual sources or by the temporal variation of a single source. We, however, note that the present broken-up spectra observed by \textit{RHESSI} pertains to the loop-top source only, and therefore does not fit into this scenario.

In the second scenario, the broken-up photon spectrum is due to physical characteristics of a certain
acceleration mechanism and/or a transport process of electrons. We consider two acceleration models as
more relevant to our observation. One is the model of diffusive shock acceleration at the flare
termination shock developed by \cite{lig13}. They suggested that electrons with a few hundred keV can
further resonate with MHD turbulence in the inertial range to achieve additional acceleration. This
results in a hardening spectrum around 500--600~keV, and the power-law indices of the electron energy distribution below and above the break energy are $\alpha_{1,2} \approx$10 and 5, respectively. After converting the model electron energy distribution to photon spectrum under the thick-target approximation of bremsstrahlung \cite{white11}, we find $\gamma_1\approx8.5$, $\gamma_2\approx3.5$, comparable with our fitting results at $t_2$.  The break energy in their model, however, amounts to 200--300~keV in the photon spectrum, much higher than our results. In their model, the break energy is associated with the spatial scale separating the inertial range and the dissipation range of turbulence, which could be either Larmor radius or inertial length of ions. The predicted break energy therefore varies with plasma temperature, density, and magnetic field. For typical coronal values of these parameters, the model predicts a break energy much higher than derived from our data analysis, with a rough estimate (not shown here).

The other model includes stochastic acceleration via wave-particle interactions in the presence of Coulomb collisions. As presented in \cite{H92}, Coulomb collisions are more effective at energies below a certain energy threshold, $E_c$, where acceleration by waves is balanced by Coulomb collisional energy loss. Thus, electron distribution below $E_c$ becomes quasi-thermal, and that above $E_c$ is non-thermal and harder. According to their results, the spectra show gradual steepening at low energies ($<30$~keV) together with decreasing $E_c$ (equivalent to $E_b$ in our study), as density of ambient plasmas increases.

Finally, we consider the third scenario related to the so-called trap-and-precipitation model, in which Fokker-Planck solutions for electrons in a magnetic trap are used to demonstrate the hardening of electron energy distribution in the trap \citep{lee00,minoshima08}. In this model, the hardening occurs because lower-energy electrons either lose energy or escape from the trap more rapidly under Coulomb collisions while higher-energy electrons can survive longer in the trap. This model describes the transport effect with acceleration mechanisms unspecified. Other than that, the same physics is already included in \cite{H92}.

\section{Conclusion}
We have presented a spectral analysis of a strong looptop HXR source during the SOL2017-09-10T16:06 flare, for which a main issue was the pileup effect on the \textit{RHESSI} spectra.
After several tests of the \textit{RHESSI} spectral fittings, we chose the pileup correction as high as $p = 2$ in order to accommodate the strong pileup effect in this event.
It follows that the loop-top source during and after the HXR maximum phase consists of broken-up power law spectrum from $\gamma=5.7$ to 3.3. The break energy of the spectra is around $E_b\approx$~46--50~keV, which tends to be lower than those reported in previous studies. The spectrum above $E_b$ becomes harder with time, and the difference of indices below and above $E_b$ increases from 1.5 to 5. It is suggested that the loop-top source may work as a thick-target to the bremsstrahlung emission due to the dense plasma in the loop-top, and that the broken-up spectrum is due to the corresponding hardening of electron energy distribution while trapped in the loop top.

We tentatively suggest that either the diffusive shock (or stochastic acceleration) or the transport effects under Coulomb collisions should be more relevant to the present observation, rather than the superposition of different sources. However, further studies are needed to figure out which of the above processes is more dominant or a combination of multiple processes is responsible for the detailed behavior of the HXR spectrum derived in this study.
This study also shows that a combination of a single power-law nonthermal component and a super-hot component cannot be ruled out, when a strong pileup effect as high as $p=2$ is assumed. In order to confirm the possibility of the double power-law spectrum in similar events, more advanced X-ray instruments of solar flares such as that aboard the \textit{Advanced Space-based Solar Observatory} \citep[\textit{ASO-S};][]{Gan15}, the Spectrometer Telescope for Imaging X-rays \citep[STIX;][]{k16} onboard the \textit{Solar Orbiter} \citep[][]{m13}, and the \textit{Focusing Optics X-ray Solar Imager} \citep[\textit{FOXSI};][]{christe17} with more events and data with higher quality will be neeeded to confirm the  broken-up HXR spectra in the looptop source.

\normalem
\begin{acknowledgements}
The authors thank the \textit{RHESSI} and \textit{SDO} teams for the high-quality X-ray and EUV data. We thank Prof. Vah{\'e} Petrosian for constructive comments on the stochastic acceleration model. We also thank Dr. Baolin Tan for helpful discussion. This work was supported by NNSFC grants 11790303 (11790300), 41774180, 11703017, and 11873036. X.K. also acknowledges the support from the Young Elite Scientists Sponsorship Program by China Association for Science and Technology, and the Young Scholars Program of Shandong University, Weihai. Y.S. acknowledges the Joint Research Fund in Astronomy (U1631242, U1731241) under the cooperative agreement between NSFC and CAS, the Major International Joint Research Project (11820101002) of NSFC and the ``Thousand Young Talents Plan''.

\end{acknowledgements}

\bibliographystyle{raa}

\end{document}